\documentclass[aps,prb,floatfix,twocolumn,showpacs,preprintnumbers,amsmath,amssymb,]{revtex4-1}

\usepackage{graphicx}
\usepackage{color}
\usepackage{amsmath}
\usepackage{amssymb}
\usepackage{amsthm}
\usepackage{stmaryrd}
\usepackage{float}

\usepackage{multirow}  
\usepackage{dcolumn}   
\usepackage{bm}        
\usepackage{stmaryrd}  
\usepackage{enumerate}
\usepackage{float}

\usepackage{hyperref}

\def  \bsig    {\mbox{\boldmath$\sigma$}}

\def  \btau    {\mbox{\boldmath$\tau$}}

\begin{document}

\preprint{}
\title{Current-induced spin polarization of a magnetized two-dimensional electron gas with Rashba spin-orbit interaction}
\author{A. Dyrda\l$^{1}$, J.~Barna\'s$^{1,2}$ and V. K.~Dugaev$^{3}$}
\address{$^1$Faculty of Physics, Adam Mickiewicz University,
ul. Umultowska 85, 61-614 Pozna\'n, Poland \\
$^2$  Institute of Molecular Physics, Polish Academy of Sciences,
ul. M. Smoluchowskiego 17, 60-179 Pozna\'n, Poland\\
$^3$ Department of Physics and Medical Engineering, Rzesz\'ow University of Technology, al. Powsta\'nc\'ow Warszawy 6, 35-959 Rzesz\'ow, Poland}
\date{\today}

\begin{abstract}
Current-induced spin polarization in a two-dimensional electron gas with Rashba spin-orbit interaction is considered theoretically in terms of the Matsubara Green functions. This formalism allows to describe temperature dependence of the induced spin polarization. The electron gas is assumed to be coupled to a magnetic substrate {\it via} exchange interaction. Analytical and numerical results on the temperature dependence of spin polarization have been obtained in the linear response regime. The spin polarization has been presented as a sum of two terms -- one proportional to the relaxation time and the other related to the Berry phase corresponding to the electronic bands of the magnetized Rashba gas. The spin-orbit torque due to Rashba interaction is also discussed. Such a torque appears as a result of the exchange coupling between the non-equilibrium spin polarization and magnetic moment of the underlayer.
\end{abstract}

\maketitle

\section{Introduction}

Spin-orbit interaction leads, in general, to a number of  interesting transport phenomena, that enable generation and control of spin currents in a pure electrical manner. Two of the most prominent examples are the spin Hall and spin Nernst effects. The former (latter) effect consists in generation of pure spin current flowing perpendicularly to an external electric field (temperature gradient) applied to the system. These effects play currently an essential role in the processes of electrical generation and detection of spin currents\cite{Sinova_RMP2015,SinovaZutic2012}. For instance, the spin current can be used as origin of spin torque exerted on magnetic moments of a ferromagnet in a bilayer system consisting of a magnetic layer attached to a nonmagnetic one with strong spin-orbit coupling. This torque, in turn,  may induce magnetic dynamics and  even can reverse magnetic moment of the magnetic layer when the spin current exceeds some critical value.

Another consequence of the spin-orbit interaction in a system with mobile electrons is the current-induced  nonequilibrium spin polarization of conduction electrons. This effect was predicted theoretically in the '70s~\cite{dyakonov71,Ivchenko78} for a two-dimensional electron gas (2DEG) with Rashba spin-orbit interaction, and then it was studied in various systems exhibiting spin-orbit interaction~\cite{edelstein90,aronov89,liu08,gorini08,wang09,schwab10,golub11,dyrdal13,dyrdal14}. The current-induced spin polarization was also observed experimentally~\cite{vorobev1979}, and currently it is attracting attention of many researchers~\cite{kato04,silov04,sih05,yang06,stern06,koehl09,kuhlen12,norman14}.

The current-induced spin polarization can also arise in a magnetic system, when it includes spin-orbit coupling. In such a case the induced non-equilibrium spin polarization interacts with the local magnetization {\it via} exchange coupling and creates a torque exerted on the magnetic moment~\cite{Manchon08,Abiague09,Gambardella11,Garello13,Kurebayashi14}.
Moreover, it has been also shown that not only external electric field, but also a temperature gradient may lead to spin-orbit driven spin polarization~\cite{dyrdal13,wang10,XiaoMa2016}. These observations initiated a wide interest  in the field- and thermally-induced spin-orbit torques and new ways of magnetization switching, that could be alternative to the switching induced by spin transfer torques~\cite{Li04,Hatami07,Ansermet10}.

In this paper we present theoretical results on the current-induced spin polarization of a magnetic 2DEG with Rashba spin-orbit interaction. Such a system is a basic model of various magnetic semiconductor heterostructures. The system consists of a 2DEG deposited on a magnetic substrate and interacting with the substrate {\it via} exchange interaction (see also Fig.1). To calculate the current-induced spin polarization we use the Matsubara Green function formalism which enables description of the temperature variation of the induced spin polarization. We derive some general formulas for the polarization and also present numerical results. The induced spin polarization is shown to include generally a term due to Berry curvature of the corresponding electron bands. Similar terms also appear in the spin-orbit torques following from exchange interaction of the electrons and magnetic underlayer.

The paper is organized as follows. In section 2 we describe the model system and also present the theoretical formalism and derive general formulas for the current-induced spin polarization.  In Section 3 we present analytical and numerical results in some specific situations; first, we consider the nonequilibrium spin polarization in the absence of  exchange field (Section 3 A), then we present results for  exchange field oriented perpendicularly to the plane of 2DEG (Section 3 B) and for exchange field  oriented in plane of 2DEG and  collinear (perpendicular) to the electric current, Section 3C (Section 3 D). In Section 4 we discuss the spin polarization in a general case of  arbitrarily oriented  exchange field. In Section 5, in turn, we consider relation of the nonequilibrium spin polarization with the Berry curvature of the corresponding electronic bands. The induced spin-orbit torque is briefly discussed in Section 6, while summary and final conclusions are in Section 7.

\section{Theoretical outline} 

We consider a magnetized 2DEG with Rashba spin-orbit interaction, as shown schematically in Fig.1. The 2DEG is assumed to be deposited on a ferromagnetic substrate which creates an effective exchange field acting on the electron gas.
\begin{figure}[t]
	\centering
	\includegraphics[width=1.00\columnwidth]{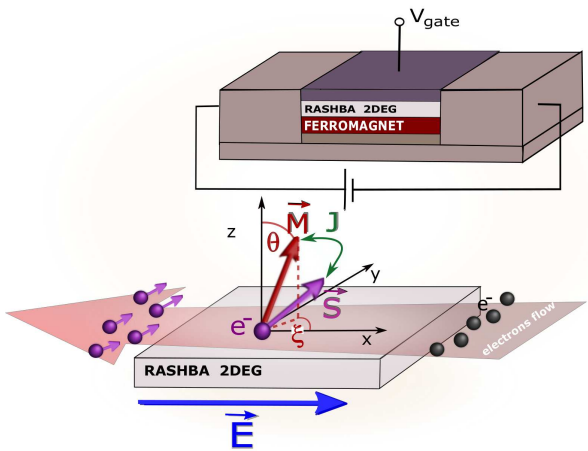}
	\caption{Current-induced spin polarization in a magnetized 2DEG. Schematic of the device (top), and coordinates used in theoretical description. Electric filed is oriented along the axis $x$.}
\end{figure}

\subsection{Model}
The single-particle Hamiltonian describing such a system can be written in the following form:
\begin{equation}
\label{H}
\hat{H} = \frac{\hbar^{2} k^{2}}{2 m} \sigma_{0} + \alpha (k_{y} \sigma_{x} - k_{x} \sigma_{y}) + \mathbf{H} \cdot \bsig ,
\end{equation}
where  $\sigma_{0}$ and $\sigma_{n}$ (for  $n = \{x, y, z\}$) are the unit and Pauli matrices   defined in the spin space, the parameter $\alpha$ in the second term of the Hamiltonian describes strength of the Rashba spin-orbit interaction, while $k_x$ and $k_y$ are the in-plane wavevector components. The third term of the above Hamiltonian describes  the effect of exchange field due to a magnetic substrate. This exchange field can be written as $\mathbf{H} =J\mathbf{M}$, with $J$ standing for the  exchange parameter ($J>0$ for a ferromagnetic coupling between the 2DEG and magnetic substrate). Note, the exchange field $\mathbf{H}$ is measured here in energy units. In spherical coordinates (see Fig.1),  components of the exchange field,  $\mathbf{H} = (H_{x}, H_{y}, H_{z})$, can be written as
\begin{subequations}
	\begin{align}
	H_{x} & = JM_x = JM\sin (\theta ) \cos (\xi ),\\
	H_{y} & = JM_y = JM\sin (\theta )\sin (\xi ),\\
	H_{z} & = JM_z = JM\cos (\theta ),
	\end{align}
\end{subequations}
where $M=|\mathbf{M}|$, while $\theta$ and $\xi$ are the polar and azimuthal angles, as defined in Fig. 1. In general, we take into account the temperature dependence of the magnetization  $M(T)$, and assume it obeys the Bloch's law
$M=M(T) = M_{0} \left[1 - \left(T/T_{c}\right)^{3/2} \right]$, where $T_c$ is the Curie temperature of the magnetic substrate, and $M_0$ is the corresponding zero-temperature magnetization.

Eigenvalues of the Hamiltonian (\ref{H}) take the form
\begin{eqnarray}
E_{\pm}
= \varepsilon_{k} \pm \lambda_{\mathbf{k}} ,
\end{eqnarray}
where $\varepsilon_{k} = \hbar^{2} k^{2}/2 m$ (with $k^{2} = k_{x}^{2} + k_{y}^{2}$), while  $\lambda_{\mathbf{k}} = [H^{2} + \alpha^{2} k^{2} -
2\alpha (H_yk_x -H_xk_y)]^{1/2}$.

Below we present the theoretical method based on the Matsubara-Green function formalism, and also derive a general formula for the nonequilibrium spin polarization induced by an external electric field.

\subsection{Method and general solution for current-induced spin polarization}

To describe spin polarization induced by an external electric field we introduce
a time-dependent external electromagnetic field of frequency $\omega/\hbar$ (note, here $\omega$ is energy) described by the vector potential
$\mathbf{A}(t)=\mathbf{A}(\omega)\exp (-i\omega t/\hbar) $.
The  electric field is  related to $\mathbf{A}$ {\it via} the formula $\mathbf{A}(\omega)=(\hbar /i\omega )\mathbf{E}(\omega)$.  Hamiltonian $H_{\mathbf{A}}^{\scriptstyle{E}}$ describing interaction of the system with the external field (treated as a perturbation) takes the form
\begin{equation}
\label{3}
\hat{H}_{\mathbf{A}}^{\scriptstyle{E}}(t) =  - \hat{\mathbf{j}}^{el}\cdot\mathbf{A}(t).
\end{equation}
Here, the operator of the electric current density is defined as $\hat{\mathbf{j}}^{el} = e \hat{\mathbf{v}}$; with  $e$ being the charge of electron ($e < 0$), and   $\hat{\mathbf{v}}=(1/\hbar )\partial \hat{H}/\partial \mathbf k $ being the electron velocity operator. The  $x$ and $y$ components of the velocity operator have the following explicit form:
\begin{eqnarray}
\label{vx}
\hat{v}_{x} = \frac{\hbar k}{m} \cos(\phi) \sigma_{0} - \frac{\alpha}{\hbar} \sigma_{y},\\
\label{vy}
\hat{v}_{y} = \frac{\hbar k}{m} \sin(\phi) \sigma_{0} + \frac{\alpha}{\hbar} \sigma_{x},
\end{eqnarray}
where $\phi$ is the angle between the wavevector $\bf k$ and the axis $x$, i.e. $k_x=k\cos (\phi)$ and $k_y=k\sin (\phi)$, while the last terms in Eq.~(\ref{vx}) and  Eq.~(\ref{vy}) represent components of the anomalous velocity that  originates from the Rashba spin-orbit interaction.

Without loss of generality, we assume in this paper that the external electric field is oriented along the $x$-axis. Thus, the $\alpha$-th ($\alpha =x, y, z$) component of the quantum-mechanical average value of  spin polarization
induced by the external electric field can be found in the Matsubara-Green functions formalism  from the following formula:
\begin{equation}
\label{SalphaE}
S_{\alpha}(i \omega_{m})= \frac{1}{\beta}  \sum_{\mathbf{k}, n} \mathrm{Tr}\left\{\hat{s}_{\alpha} G_{\mathbf{k}}(i \varepsilon_{n} + i \omega_{m}) \hat{H}_{\mathbf{A}}^{\scriptstyle{E}} (i \omega_{m})G_{\mathbf{k}}(i \varepsilon_{n}) \right\},
\end{equation}
where $\hat{s}_{\alpha} = \hbar \sigma_{\alpha}/2$ is the operator of the $\alpha$'s spin component,  $\beta =1/k_BT$ (with $T$ and $k_B$ denoting the temperature and Boltzmann constant, respectively), $\varepsilon_{n}=(2n+1)i\pi k_BT$ and $\omega_{m}=2mi\pi k_BT$ are the Matsubara energies, while  $G_{\mathbf{k}}(i \varepsilon_{n})$ are the Matsubara Green functions (in the $2\times 2$ matrix form).
Note, the  perturbation term takes now the form $\hat{H}_{\mathbf{A}}^{\scriptstyle{E}}(i \omega_{m})=-e\hat{v}_xA_x(i \omega_{m})$, with the amplitude of the vector potential $A_{x}(i \omega_{m})$ determined by the amplitude $E_x(i \omega_{m})$ of electric field through the relation $A_{x}(i \omega_{m}) = \frac{E_{x}(i \omega_{m}) \hbar}{i (i \omega_{m})}$.

Taking into account the explicit form of $\hat{H}_{\mathbf{A}}^{\scriptstyle{E}}(i \omega_{m})$,
one can rewrite Eq.(\ref{SalphaE})  in the form
\begin{eqnarray}
\label{4}
S_{\alpha} (i \omega_{m})= - \frac{1}{\beta}\frac{e E_{x}(i\omega_m) \hbar}{i (i \omega_{m})}
\hspace{3.6cm} \nonumber \\
\times \sum_{\mathbf{k}, n} \mathrm{Tr}\left\{\hat{s}_{\alpha} G_{\mathbf{k}}(i \varepsilon_{n} + i \omega_{m}) \hat{v}_{x} G_{\mathbf{k}}(i \varepsilon_{n}) \right\}.
\end{eqnarray}
The sum over Matsubara energies in the above expression can be
calculated by the method of contour integration, \cite{abrikosov,mahan}
\begin{eqnarray}
\label{5}
\frac{1}{\beta} \sum_{n} \hat{s}_{\alpha} G_{\mathbf{k}}(i \varepsilon_{n} + i \omega_{m}) \hat{v}_{x} G_{\mathbf{k}}(i \varepsilon_{n}) \hspace{0.8cm} \nonumber\\
= - \int_{\mathcal{C}}  \frac{dz}{2 \pi i} f(z) \hat{s}_{\alpha} G_{\mathbf{k}}(z + i \omega_{m}) \hat{v}_{x} G_{\mathbf{k}}(z) ,
\end{eqnarray}
where $\mathcal{C}$ denotes the appropriate contour of integration and $f(z)$ is a meromorphic function of the form $ ({\mathrm{e}}^{\beta z} + 1)^{-1}$, that has simple poles at the odd integers $n$, $z = i \varepsilon_{n}$ (for details see Refs~[\onlinecite{abrikosov,mahan}]).

Upon analytical continuation one obtains
\begin{eqnarray}
\label{SalphaE_final}
S_{\alpha}(\omega) = \hspace{7.2cm}\nonumber \\
- \frac{e \hbar}{\omega} E_{x} {\mathrm{Tr}} \sum_{\mathbf{k}} \int \frac{d \varepsilon}{2 \pi} f(\varepsilon) \hat{s}_{\alpha} \Bigl( G_{\mathbf{k}}^{R}(\varepsilon + \omega) \hat{v}_{x} [G_{\mathbf{k}}^{R}(\varepsilon) - G_{\mathbf{k}}^{A}(\varepsilon)]\Bigr. \nonumber\\
+ \Bigl. [G_{\mathbf{k}}^{R}(\varepsilon) - G_{\mathbf{k}}^{A}(\varepsilon)] \hat{v}_{x} G_{\mathbf{k}}^{A}(\varepsilon - \omega)\Bigr).\hspace{1cm}
\end{eqnarray}
Here, $f(\varepsilon )$ is the Fermi-Dirac distribution function and  $G_{\mathbf{k}}^{R/A}(\varepsilon)$  is the impurity-averaged retarded/advanced Green function corresponding to the Hamiltonian (1). The Green functions take the following explicit form:
\begin{eqnarray}
G_{\mathbf{k}}^{R/A}(\varepsilon) = G_{\mathbf{k}\, 0}^{R/A}(\varepsilon) \sigma_{0}\hspace{4.5cm}\nonumber\\ + G_{\mathbf{k}\, x}^{R/A}(\varepsilon) \sigma_{x} + G_{\mathbf{k}\, y}^{R/A} (\varepsilon) \sigma_{y} + G_{\mathbf{k}\, z}^{R/A}(\varepsilon) \sigma_{z}, \hspace{0.4cm}
\end{eqnarray}
where
\begin{subequations}
	\begin{align}
	G_{\mathbf{k}\, 0}^{R/A}(\varepsilon) &= \frac{1}{2} [G_{+}(\varepsilon) + G_{-}(\varepsilon)], \\
	G_{\mathbf{k}\, x}^{R/A}(\varepsilon) &= \frac{1}{2 \lambda_{\mathbf{k}}} (\alpha k_{y} + H_x)[G_{+}(\varepsilon) -G_{-}(\varepsilon)],\\
	G_{\mathbf{k}\, y}^{R/A}(\varepsilon) &= \frac{1}{2 \lambda_{\mathbf{k}}} (-\alpha k_{x} + H_y)[G_{+}(\varepsilon)-G_{-}(\varepsilon)],\\
	G_{\mathbf{k}\, z}^{R/A}(\varepsilon) &=  \frac{1}{2 \lambda_{\mathbf{k}}} H_z [G_{+}(\varepsilon) -G_{-}(\varepsilon)],
	\end{align}
\end{subequations}
with $G_{\pm}^{R}(\varepsilon) = [\varepsilon + \mu - E_{\pm}  + i\Gamma]^{-1}$ and $G_{\pm}^{A}(\varepsilon) = [\varepsilon + \mu - E_{\pm}  - i\Gamma]^{-1}$. Note, we assumed  $\Gamma = \hbar/2\tau$, with equal effective relaxation time $\tau$ in the two subbands.

 Using equation (\ref{SalphaE_final}) as a starting point and performing integration over $\varepsilon$ we get finally  the following formula for the three components of the current-induced spin polarization:
 \begin{eqnarray}\label{S_x}
 S_{x} = - e E_{x} \hbar \hspace{6.3cm}\nonumber\\
 \times\int \frac{d^{2} \mathbf{k}}{(2\pi)^{2}} \left\{\frac{1}{2 \Gamma} \frac{\hbar^{2} k_{x}}{2 m \lambda_{\mathbf{k}}} (\alpha k_{y} + H_{x}) [f'(E_{+}) - f'(E_{-})] \right. \nonumber\\
 + \frac{\alpha}{\Gamma} \frac{(\alpha k_{y} + H_{x}) (\alpha k_{x} - H_{y})}{(2\lambda_{\mathbf{k}})^{2} + (2\Gamma)^{2}} [f'(E_{+}) + f'(E_{-})] \nonumber\\
 - \frac{ \alpha H_{z}}{(2 \lambda_{\mathbf{k}})^{2}} \frac{(2 \Gamma)^{2}}{(2\lambda_{\mathbf{k}})^{2} + (2\Gamma)^{2}} [f'(E_{+}) + f'(E_{-})] \nonumber\\
 - \left. \frac{\alpha H_{z}}{4 \lambda_{\mathbf{k}}^{3}} [f(E_{-}) - f(E_{+})]\right\},\hspace{0.6cm}
 \end{eqnarray}
 \begin{eqnarray}\label{S_y}
 S_{y} = e E_{x} \hbar \hspace{7cm}\nonumber\\
 \times\int \frac{d^{2} \mathbf{k}}{(2\pi)^{2}} \left\{ \frac{1}{4\Gamma} \frac{\alpha}{\lambda_{\mathbf{k}}^{2}} (\alpha k_{x} - H_{y})^{2} [f'(E_{+}) + f'(E_{-})]\right.\hspace{0.5cm}\nonumber\\
 +\frac{\hbar^{2}}{m \lambda_{\mathbf{k}}} (\alpha k_{x} - H_{y}) \frac{1}{4\Gamma} [f'(E_{+}) - f'(E_{-})]\hspace{2.3cm} \nonumber\\
 + \left. \frac{\alpha \Gamma}{\lambda_{\mathbf{k}}^{2}} \left(1 - \frac{(\alpha k_{x} - H_{y})^{2}}{\lambda_{\mathbf{k}}^{2}}\right) \frac{f'(E_{+}) + f'(E_{-})}{(E_{+} - E_{-})^{2} + (2\Gamma)^{2}}\right\}, \hspace{0.8cm}
 \end{eqnarray}
 \begin{eqnarray}\label{S_z}
 S_{z} = - e E_{x} \hbar \hspace{7.0cm}\nonumber\\
 \times \int \frac{d^{2} \mathbf{k}}{(2\pi)^{2}} \left\{  \frac{ \alpha H_{z}}{\Gamma} \frac{\alpha k_{x} - H_{y}}{(2 \lambda_{\mathbf{k}})^{2} + (2\Gamma)^{2}} [f'(E_{+}) + f'(E_{-})] \right.\hspace{0.5cm}\nonumber\\
 + \frac{\hbar^{2} k_{x}}{2m} \frac{H_{z}}{\lambda_{\mathbf{k}}} \frac{1}{2\Gamma} [f'(E_{+}) - f'(E_{-})]\hspace{1.8cm}\nonumber\\
 - \frac{\alpha}{4 \lambda_{\mathbf{k}}^{2}} \frac{(2\Gamma)^{2} (H_{x} + \alpha k_{y})}{(2 \lambda_{\mathbf{k}})^{2} + (2\Gamma)^{2}} [f'(E_{+}) + f'(E_{-})]\hspace{0.3cm}\nonumber\\
 -\left. \frac{\alpha}{4 \lambda_{\mathbf{k}}^{3}} (H_{x} + \alpha k_{y}) [f(E_{+}) - f(E_{-})]\right\}.\hspace{0.8cm}
 \end{eqnarray}
 Details on the derivation of the  above equations are presented in the Appendix A.
 Before presenting results on the current-induced spin polarization for an arbitrary orientation of the exchange field, we consider first some special cases.

\section{Special cases} 

\subsection{Zero exchange field}

First, we reconsider the limit of zero exchange field, i.e. the limit of a nonmagnetized 2DEG, when only the $y$ component of spin polarization survives. The general expression for $S_{y}$ takes then the following form:
\begin{eqnarray}
S_{y} = e E_{x} \hbar \int \frac{dk k}{(2\pi)^{2}} \left\{ \alpha \frac{\pi}{4 \Gamma} [f'(E_{+}) + f'(E_{-})] \right. \nonumber\\
+ \alpha \pi \Gamma \frac{f'(E_{+}) + f'(E_{-})}{(2\alpha k)^{2} + (2\Gamma)^{2}}\nonumber\\
+ \left. \pi \frac{\hbar^{2} k}{4 m \Gamma} [f'(E_{+}) - f'(E_{-})]  \right\},
\end{eqnarray}
where $ f'=\partial f/\partial E$.
Note that in this limit the eigenvalues have the form $E_{\pm} = \varepsilon_k \pm \alpha k$.

In the low-temperature regime, the above integrals can be evaluated analytically and one arrives at
\begin{eqnarray}
S_{y} = - e E_{x} \hbar \int \frac{dk k}{(2\pi)^{2}} \left\{ \alpha \frac{\pi}{4 \Gamma} [\delta(E_{+}-\mu) + \delta(E_{-}-\mu)] \right. \nonumber\\
+ \alpha \pi \Gamma \frac{\delta(E_{+}-\mu) + \delta(E_{-}-\mu)}{(2\alpha k)^{2} + (2\Gamma)^{2}}\nonumber\\
+ \left. \pi \frac{\hbar^{2} k}{4 m \Gamma} [\delta(E_{+}-\mu) - \delta(E_{-}-\mu)]  \right\}.\hspace{0.5cm}
\end{eqnarray}
When both subbands are occupied (which corresponds to $\mu > 0$), the Dirac delta functions in the above equation can be written in the form
\begin{equation}\label{DiracDelta}
\delta(E_{\pm} - \mu) = \frac{m}{\sqrt{2 m\mu \hbar^{2} + m^{2} \alpha^{2}}} \delta(k - k_{\pm}),
\end{equation}
and finally one obtains
\begin{eqnarray}
\label{Sy_Edelsteinlike}
S_{y} = \frac{1}{2} e E_{x} \frac{m \alpha}{2\pi \hbar^{2}} \tau\hspace{5.0cm}\nonumber\\
- e E_{x} \frac{\hbar m \alpha}{16 \pi \Gamma \sqrt{2 m \mu \hbar^{2} + m^{2} \alpha^{2}} }\hspace{2.9cm}\nonumber\\
\times
 \left[ \frac{k_{+}}{1 + (\alpha k_{+}/\Gamma)^{2}} + \frac{k_{-}}{1 + (\alpha k_{-}/\Gamma)^{2}}\right],\hspace{0.5cm}
\end{eqnarray}
with $k_{\pm} = \mp \frac{m \alpha}{\hbar^{2}} + \frac{1}{\hbar^{2}} \sqrt{m^{2} \alpha^{2} + 2 m \mu \hbar^{2}}$.
The first term of Eq.(\ref{Sy_Edelsteinlike}) corresponds to the Edelstein expression for the current-induced spin polarization in the so-called {\it bubble approximation},
\begin{equation}
S^0_{y} = \frac{1}{2} e E_{x} \frac{m \alpha}{2\pi \hbar^{2}} \tau .
\end{equation}
Note, the impurity vertex correction is neglected in our  considerations. Such a correction leads to some renormalization of the spin polarization (for details see e.g. Ref. [\onlinecite{edelstein90,BDDIchapt}]). The second term in (\ref{Sy_Edelsteinlike}) is a correction which originates from the imaginary  term in the nominator of the Green function and  products of two retarded or two advanced Green's functions (omitted in Ref.~\onlinecite{edelstein90}). Note, the second term in Eq.(\ref{Sy_Edelsteinlike}) vanishes in the quasi-ballistic limit (low impurities concentration), when $\Gamma\to 0$.

\begin{figure}[]
	\centering
	\includegraphics[width=\columnwidth]{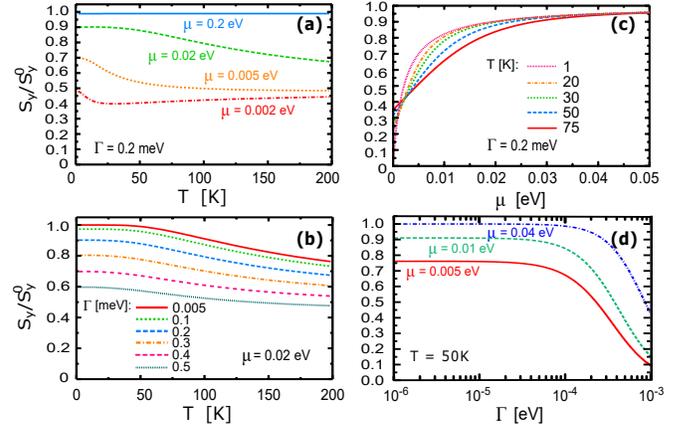}
	\caption{Temperature dependence of the spin polarization  for different values of $\mu$ (a) and for different values of $\Gamma$ (b). Spin polarization as a function of chemical potential for different  values of temperature $T$ (c), and spin polarization as a function of $\Gamma$ for different values of chemical potential $\mu$ (d). The spin polarization is normalized to the Edelstein term $S_y^0$ [see Eq.(20)]. In numerical calculations it was assumed: $m = 0.07 m_{0}$ ($m_{0}$ being the electron rest mass), $\alpha = 2 \cdot 10^{-11}$eVm, and $eE_x = - 5 \cdot 10^{-2}$ eV/m. Other parameters as indicated.}
	\label{zero}
\end{figure}

In the general case, i.e. for arbitrary $T$ and arbitrary chemical potential $\mu$, one should use the general formula (16). However, one point requires some comment. It is known, that for impurities with short-range ($\delta$-like) potential and $\mu>0$, the parameter $\Gamma$ is constant, while for negative $\mu$ it increases and diverges when
$\mu$ approaches the bottom of the lower energy band~\cite{Brosco,Dyrdal2016}. Thus, at a certain value of $\mu$,  $\mu =\mu_{\rm loc}$, the Ioffe-Regel localization condition~\cite{Ioffe} is obeyed, and the states become localized below $\mu_{\rm loc}$. Accordingly, the results are valid beyond the localization regime, i.e. for $\mu >\mu_{\rm loc}$.

Now, we present some numerical results. In Fig.\ref{zero}(a) we show the temperature dependence of spin polarization for four different values of chemical potential $\mu$. Here, we should mention that the chemical potential also depends on temperature, thus a fixed value of chemical potential means that the carrier concentration varies. If however the system is gated one can keep chemical potential constant. Apart from this, the relaxation time $\tau$ (and thus the parameter $\Gamma$) may also depend on temperature $T$. This dependence, however, is neglected in Fig.\ref{zero}. The spin polarization $S_y$ was obtained from Eq.(16) and is normalized there to the corresponding value of $S^0_y$ (note $S^0_y$ does not depend explicitly on temperature). For the largest value of $\mu$, the $S_y$ component remains almost constant in the temperature range shown in Fig.\ref{zero}(a), and is roughly equal to the corresponding value of $S^0_y$. For smaller values of $\mu$,  in turn, the spin polarization becomes reduced monotonously with increasing $T$ (see the curves for $\mu =0.02$ eV and $\mu = 0.005$ eV). For still lower values of $\mu$, the temperature dependence is nonmonotonous - it first decreases and then slightly increases with temperature. To understand this behaviour we plot in Fig.\ref{zero}(c) the spin polarization $S_y$  as a function of chemical potential for several values of temperature and the same $\Gamma$ as in Fig.\ref{zero}(a).  This figure clearly shows that spin polarization tends to $S^0_y$ with increasing $\mu$. Such a behavior is reasonable as the second term in Eq.(19) decreases with increasing $\mu$ (the effective role of finite $\Gamma$ decreases with increasing $E_{+} - E_{-}$). For small values of $\mu$, however, the second term in Eq.(19) plays a role and the spin polarization is reduced. The temperature dependence appears when $E_{+} - E_{-}$  at the Fermi level is of the order or smaller than $kT$, which takes place in the region of small values of $\mu$. Moreover, this figure also shows that $S_y$ decreases with increasing $T$, except a narrow region of small values of $\mu$, where the temperature dependence  is nonmonotonous, exactly like in Fig.\ref{zero}(a)
The temperature dependence is also shown in Fig.\ref{zero}(b) for several values of $\Gamma$. This figure also shows that the correction due to the second term in Eq.(19) increases with increasing $\Gamma$. The latter behavior is shown explicitly in  Fig.\ref{zero}(d) for indicated  values of $\mu$. The decrease of spin polarization with increasing $\Gamma$ is physically clear as the effective separation of the two Rashba bands becomes reduced with increasing $\Gamma$. The second term in Eq.(19) plays then an important role and leads to reduction of spin polarization.

\subsection{Exchange field perpendicular to plane of 2DEG}

Consider now a magnetized 2DEG and let us begin with the situation when the exchange field (or equivalently substrate magnetization) is perpendicular to the plane of 2DEG, $\mathbf{H} = (0, 0, H_{z})$.
The eigenvalues of Hamiltonian (1) reduce now to the form $E_{\pm} = \varepsilon_{k} \pm \zeta$, with $\zeta = \sqrt{H^{2} + \alpha^{2} k^{2}}$.

\begin{figure*}[]
	\centering
	\includegraphics[width=1.85\columnwidth]{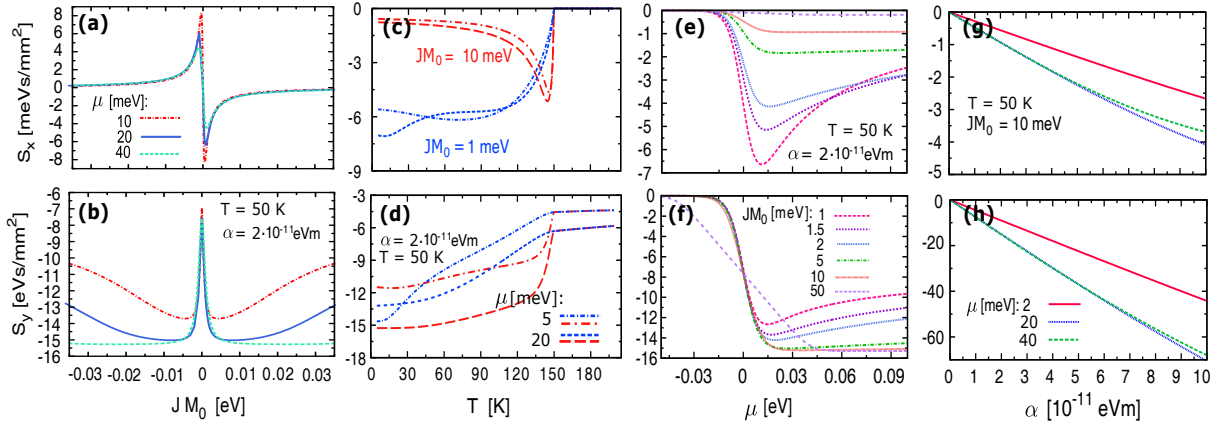}
	\caption{In-plane components of the current-induced spin polarization for the exchange field normal to the plane of 2DEG.
		(a,b) Variation of $S_x$ (a) and $S_y$ (b) components of the current-induced spin polarization with the exchange field $JM_0$ for indicated values of the chemical potential. (c,d) Dependence of $S_x$ (c) and $S_y$ (d)  on temperature for different values of magnetization $JM_0$ and chemical potential $\mu$. Here, the red (blue) curves correspond to $JM_0=10$ meV ($JM_0=1$ meV).  (e,f) The components $S_x$ (e) and $S_y$ (f) as a function of the  chemical potential for indicated values of $JM_0$. (g,h) The components $S_x$ (g) and $S_y$  (h) as a function of the Rashba coupling parameter $\alpha$ for indicated values of the chemical potential. Numerical results have been obtained for $\Gamma = 0.5\cdot10^{-6}$ eV, $T_c=150$ K, and parameters indicated in the figures (note, the parameters in the bottom panel are the same as in the top panel). The other parameters are  as in Fig.2. }
	\label{Hz_figure}
\end{figure*}

The general expressions describing the two nonzero components of spin polarization take the forms
\begin{eqnarray}\label{SxHz}
S_{x} =  e E_{x} \hbar \int \frac{dk k}{2 \pi} \frac{\alpha H_{z}}{(2 \zeta)^{2}} \left\{  \frac{\Gamma^{2}}{\zeta^{2} + \Gamma^{2}} [f'(E_{+}) - f'(E_{-})] \right.\nonumber\\
- \left. \frac{1}{\zeta} [f(E_{+}) - f(E_{-})] \right\},\hspace{0.7cm}
\end{eqnarray}
\begin{eqnarray}\label{SyHz}
S_{y} = e E_{x} \hbar \int \frac{dk k}{(2\pi)^{2}} \alpha \pi \left\{  \frac{\alpha^{2} k^{2}}{4 \Gamma \zeta^{2}} [f'(E_{+}) + f'(E_{-})] \right.\hspace{0.9cm}\nonumber\\
+ \frac{\hbar^{2} k^{2}}{4 \Gamma m \zeta} [f'(E_{+}) - f'(E_{-})] \nonumber\\
\left. + \Gamma \left(2 - \frac{\alpha^{2} k^{2}}{\zeta^{2}}\right) \frac{f'(E_{+}) + f'(E_{-})}{(2\zeta)^{2} + (2\Gamma)^{2}} \right\},\hspace{0.6cm}
\end{eqnarray}
while $S_{z} = 0$.
Accordingly, the  electric field generates now spin polarization with both in-plane components nonzero, while the component normal to the plane of 2DEG (along the exchange field) vanishes exactly. Thus, the exchange field generates spin polarization along the electric field and also modifies the spin polarization  along the axis $y$.

In the  low temperature limit equations (\ref{SxHz}) and (\ref{SyHz}) lead to the following analytical expressions:
\begin{eqnarray}\label{SxHfinal}
S_{x} = -\hbar \frac{e}{8 \pi} E_{x} \frac{H_{z}}{\alpha} \left[\frac{\zeta_{+} - \zeta_{-}}{\zeta_{+} \zeta_{-}} \right.\hspace{3.2cm}\nonumber\\
+\left.  \alpha^{2} \left(\frac{\nu_{+}}{\zeta_{+}^{2}} \frac{1}{1 + (\zeta_{+}/\Gamma)^{2}} - \frac{\nu_{-}}{\zeta_{-}^{2}} \frac{1}{1 + (\zeta_{-}/\Gamma)^{2}}\right) \right],
\end{eqnarray}
\begin{eqnarray}\label{SyHzanal}
S_{y} = - \hbar \frac{e}{16 \pi} E_{x} \frac{\alpha}{\Gamma} \left[ \frac{k_{+}^{2}}{\zeta_{+}} - \frac{k_{-}^{2}}{\zeta_{-}} \right]\hspace{2.3cm}\nonumber\\
- \hbar \frac{e}{16 \pi} E_{x} \left[\left(1-\frac{H_{z}}{\zeta_{+}} \right) \frac{\nu_{+}}{1 + (\zeta_{+}/\Gamma)^{2}}\right.\nonumber\\
\left. + \left(1-\frac{H_{z}}{\zeta_{-}} \right) \frac{\nu_{-}}{1 + (\zeta_{-}/\Gamma)^{2}} \right],
\end{eqnarray}
where $\nu_{\pm} =\frac{m}{\hbar^{2}} (1 \pm \frac{m \alpha^{2}}{\hbar^{2} \zeta_{\pm}})^{-1}$ represent  the density
 of states corresponding to the $E_{\pm}$ subbands, respectively,  $\zeta_{\pm} = \zeta (k = k_{\pm})$, and  $k_{\pm} = \frac{\sqrt{2m}}{\hbar^{2}} \sqrt{m \alpha^{2} + \mu \hbar^{2} \mp \sqrt{m^{2} \alpha^{4} + 2 m \alpha^{2} \hbar^{2} \mu + H_{z}^{2} \hbar^{4}}}$ are the Fermi wavevectors corresponding to the two subbands.

In the ballistic limit (extremely long relaxation time), Eq. (\ref{SxHz}) takes the form
\begin{eqnarray}\label{SxHz1}
S_{x} =  - e E_{x} \hbar \int \frac{dk k}{2 \pi} \frac{\alpha H_{z}}{4 \zeta^{3}}
 [f(E_{+}) - f(E_{-})],
\end{eqnarray}
which after integration over $k$ leads to
\begin{equation}\label{SxHz_ballistic}
S_{x} = - \hbar \frac{e}{8 \pi} E_{x} \frac{H_{z}}{\alpha} \frac{\zeta_{+} - \zeta_{-}}{\zeta_{+} \zeta_{-}},
\end{equation}
i.e. to the first term in Eq.(\ref{SxHfinal}).
The above expression does not depend on the relaxation time and due to the mathematical form of Eq.(\ref{SxHz1})  it may be identified as the Berry phase related contribution to the spin polarization, that in turn may be responsible for anti-damping spin-torque. For details see Section V.

Numerical results on the current-induced spin polarization of the magnetized 2DEG with the exchange field oriented perpendicularly to the plane are shown in Fig.\ref{Hz_figure}.
The dependence of $S_{x}$ and $S_{z}$ on the exchange field $JM_0$  is presented in Figs~\ref{Hz_figure}(a) and \ref{Hz_figure}(b), respectively. Figure \ref{Hz_figure}(a) clearly shows that $S_x$ vanishes in the limit of
zero exchange field and then its magnitude grows rather fast with increasing $JM_0$. Then, it decreases to zero for large exchange fields. Magnitude of $S_y$, in turn, is nozero for zero exchange field, and increases with increasing $JM_0$. It reaches a maximum at some value of $JM_0$, and then decreases with a further increase in $JM_0$. Such a behavior with $JM_0$ can be understood since the relative role of Rashba coupling decreases with increasing $JM_0$. Note, that the $S_x$ component is antisymmetrical with respect to sign reversal of $JM_0$, while the $S_y$ component is then symmetrical.
Figures \ref{Hz_figure}(c) and \ref{Hz_figure}(d) present the $x$ and $y$ components of spin polarization as a function of temperature for fixed values of chemical potential and $JM_0$.  In numerical calculations we have assumed $T_{c} = 150$ K and therefore the $S_{x}$ component vanishes for $T \geq 150$ K. In turn, the $S_{y}$ components is remarkably enhanced below $T_{c}$ and drops to a weakly temperature dependent value (for fixed chemical potential and the parameter $\Gamma$) when $T \geq 150$ K.

Variation of spin polarization with the chemical potential is presented in Figs~\ref{Hz_figure}(e) and \ref{Hz_figure}(f) for different magnitudes of the exchange field $JM_0$. Magnitudes of both components increase monotonously with $\mu$ when $\mu$ is inside the energy region between the bottom edges of the two subbands. For $\mu$ in the vicinity of the bottom of higher energy band, these components reach maximum values and for larger $\mu$ they decrease with increasing $\mu$. Note, the component $S_x$ is roughly three orders of magnitude smaller than the $S_y$ component.

In Fig. \ref{Hz_figure}(g) and \ref{Hz_figure}(h) we show the $x$ and $y$ components of the spin polarization as a function of the Rashba coupling constant. These figures clearly show that the absolute values of both components increase roughly linearly with $\alpha$. However, some deviations from this linear dependence appear above certain values of $\alpha$, where the increase is smaller.

\subsection{Exchange field in plane of 2DEG and perpendicular to electric field}
\begin{figure}[h!]
	\centering
	\includegraphics[width=\columnwidth]{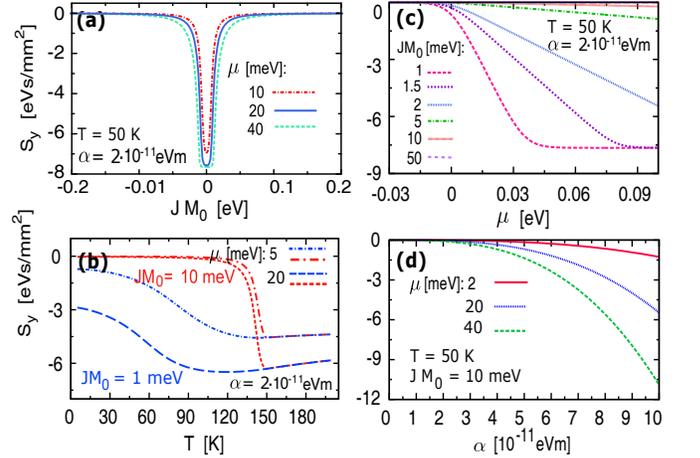}
	\caption{Current-induced spin polarization in the case when exchange field is along the $y$-axis. (a) Spin polarization   plotted as a function of the exchange field $JM_0$ for several values of chemical potential. (b) $S_y$ as a function of temperature for indicated values of $\mu$ and $JM_0$. Here, the red (blue) curves correspond to $JM_0=10$ meV ($JM_0=1$ meV). (c) Variation of the spin polarization with the chemical potential $\mu$ for fixed values of exchange field;  (d) $S_y$ as a function of the Rashba constant $\alpha$ for indicated values of the chemical potential.  Other parameters as in Fig.3.}
	\label{Hy_figure}
\end{figure}

In this section we consider the current-induced spin polarization for the magnetization vector (exchange field) oriented along  the $y$ axis, i.e. when the exchange field is in the plane of two-dimensional electron gas and perpendicular to the current. In such a case the $x$ and $z$ components of the current-induced spin polarization vanish exactly, and the only nonzero component is  $S_{y}$  - like in the case of zero exchange field. This component, however, is modified by the exchange field.

Numerical results for $S_y$ are shown in Fig.\ref{Hy_figure}, where variation of $S_y$ with the exchange field $JM_0$, Fig.\ref{Hy_figure}(a), clearly shows that the spin polarization decreases relatively fast with increasing absolute value of $JM_0$ and is suppressed when the Zeeman-like term (due to exchange coupling to the substrate) dominates over the Rashba term. The suppression to zero of spin polarization at large $JM_0$ appears due to strong modification of electronic states by the Zeeman like term, and takes place for all values of chemical potential.

Temperature dependence of $S_y$ is shown in Fig.\ref{Hy_figure}(b) for two values of chemical potential and two values of $JM_0$. For the larger value of  $JM_0$, the spin polarization $S_y$ vanishes in a broad temperature region and then increases when $T$ approaches the Curie temperature, reaching the magnitude of $S_y$ in the limit of a nonmagnetized  2DEG. This behavior is consistent with that in Fig.4a

In  Fig.\ref{Hy_figure}(c) we show $S_y$ as a function of chemical potential.
As follows from this figure, $S_y$ increases monotonously with the chemical potential increasing from the minimum of the lower subband, and then becomes saturated for large values of $\mu$. The rate of this increase as well as the  chemical potential at which the saturation appears depend on $JM_0$.
Spin polarization as a function of the Rashba parameter $\alpha$ is shown in  Fig.\ref{Hy_figure} (d) for indicated values of the exchange field. In general, the $y$ component of spin polarization increases now nonlinearly with the Rashba constant.

\subsection{Exchange field in plane of 2DEG and collinear with electric field}

\begin{figure*}[t]
	\centering
	\includegraphics[width=1.895\columnwidth]{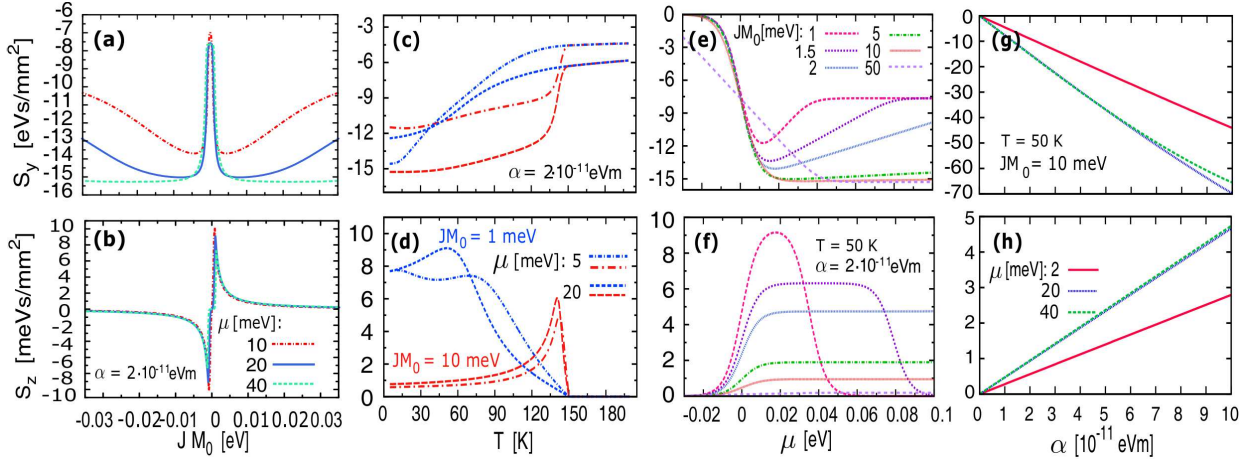}
	\caption{Current-induced spin polarization in case when the exchange field is parallel to the electric field for indicated parameters.  Variation of the spin polarization components with the exchange field for indicated values of chemical potential (a,b); with temperature for $JM_0=10$ meV (red curves),  $JM_0=1$ meV (blue curves), and for indicated chemical potential $\mu$ (c,d); with the chemical potential for indicated values of $JM_0$ (e,f); and with the Rashba coupling parameter $\alpha$ for indicated values of chemical potential (f,g). Note, the parameters used for the top panel are the same as the corresponding ones in the bottom panel. Other parameters as in Fig. 3}
	\label{Hx_figure}
\end{figure*}

When the exchange field is oriented along the $x$ axis, i.e. it is collinear with the external electric field, the $x$ component of spin polarization vanishes, whereas the $y$ and $z$ components are non-zero. In general, the $S_{z}$ component of spin polarization is roughly three orders of magnitude smaller than the $S_{y}$ component. Variation of both components with the exchange field $JM_0$, temperature, chemical potential, and Rashba constant is presented in Figs\ref{Hx_figure}~(a-d).

Behavior of the $S_y$ and $S_z$ components with $JM_0$, $T$, $\mu$ and $\alpha$ is qualitatively similar to the corresponding behavior of the components $S_y$ and $S_x$ in the case with the exchange field normal to the plane of 2DEG, see Fig.\ref{Hz_figure}. There are some differences of rather quantitative character, which follow from different electronic bands in these two situations. For instance, the $S_z$ component varies with the chemical potential in a slightly different manner than the $S_x$ component in Fig.3.
Weak difference also appear in the variation of the $S_y$ component with temperature for $T$ below the Curie temperature $T_c$. Similarly as in Fig.~3, both components behave  almost linearly with the Rashba parameter $\alpha$.

\section{Numerical results for arbitrarily oriented exchange field }

\begin{figure}[t]
	\centering
	\includegraphics[width=\columnwidth]{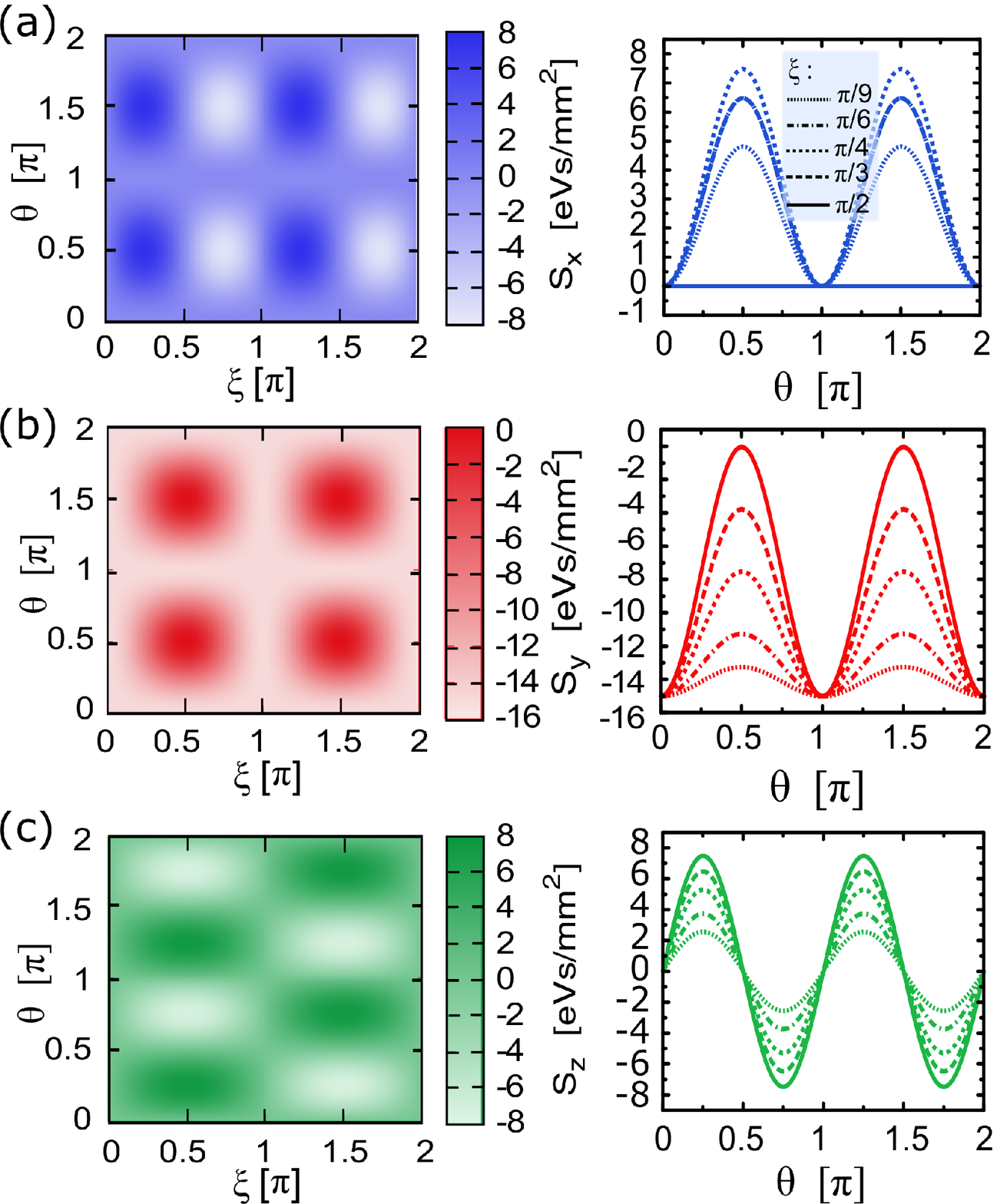}
	\caption{Spin polarization induced by an external electric field in the presence of arbitrarily oriented exchange field. Left panel presents the three components of spin polarization as a function of $\theta$ and $\xi$. The right panel shows the corresponding vertical cross-sections (for a few values of the  angle $\xi$).  The parameters assumed in numerical calculations are: $\mu = 0.02$ eV, $\alpha = 2 \cdot 10^{-11}$eVm, $JM_{0} = 0.01$eV, $\Gamma = 0.5\cdot10^{-6}$eV and $T = 50$ K. The other parameters as in Fig.3.    }
	\label{CISPEodTheta_fig}
\end{figure}

Up to now we have discussed only some specific situations, when the exchange field is oriented along the three main directions: (i) along the electric field, (ii) normal to the electric field and to the plane of 2DEG, and (iii) normal to the electric field and oriented in the plane of 2DEG. Now let us consider a general case, when the exchange field is oriented arbitrarily. This orientation is described by the polar $\theta$ and azimuthal $\xi$  angles, as shown in Fig.1. Generally, all three components of spin polarization (i.e. $S_x$, $S_y$ and $S_z$) can be nonzero.
In Fig. \ref{CISPEodTheta_fig} we present these components  as a function of both $\theta$ and $\xi$ angles, see left panel in this figure. The right panel, in turn, presents several vertical cross-sections of the corresponding density plots from the left panel. In the specific configurations  considered in the preceding section, the results shown in Fig.~6 reduce to the corresponding ones discussed in Sec.3. This figure shows the regions in the ($\theta ,\xi$) plane, where particular components of the spin polarization are large and where are small or suppressed to zero .

The results in a general case, like those presented in Fig. \ref{CISPEodTheta_fig} are required  when considering magnetic dynamics induced by spin torque due to spin polarization. Magnetic moment (and thus also exchange field) precesses then in space, and this time evolution is associated with time evolution of the spin polarization. In this paper, however,  we do not consider dynamical properties and focus rather on evaluating spin polarization in static situations.

\section{Relation with the Berry Curvature}
Recently  H. Kurebayashi et al.~\cite{Kurebayashi}, based on experimental results, have proposed the  anti-damping spin-orbit torque mediated by the Berry phase~\cite{Berry}. In other words, they showed that the Berry curvature gives rise to the spin-orbit torque in systems with broken inversion symmetry.

Our results given by Eqs. (\ref{S_x})-(\ref{S_z}) show that when the exchange field is nonzero, the inversion symmetry is broken and the general expressions for the $x$ and $z$ components of the spin polarization  contain terms that do not depend on relaxation rate, but are functions of the Fermi-Dirac distribution function instead of its derivative. Thus,  taking into account  the notation well known in the context of the anomalous Hall effect, we can rewrite Eqs~(\ref{S_x})-(\ref{S_z}) as follows:
\begin{equation}
S_{\alpha} = S_{\alpha}^{I} + S_{\alpha}^{II}
\end{equation}
where the first term depends on the states in a close vicinity of the Fermi level: $S_{\alpha}^{I} = S_{\alpha}[f'(E_{\pm})]$, while the second term contains information from all electronic states: $S_{\alpha}^{II} = S_{\alpha}[f(E_{\pm})]$.
Now we show that the terms $S_{\alpha}^{II}$ are related to the Berry curvature.

To do this let us rewrite the Hamiltonian (\ref{H}) in the following form:
\begin{equation}
H = \varepsilon_{k} \sigma_{0} + \mathbf{n} \cdot \bsig
\end{equation}
where  $\mathbf{n} = \left(\alpha k_{y} - H_{x}, - \alpha k_{x} - H_{y}, - H_{z}  \right)$, and $\varepsilon_{k} = \hbar^{2} (k_{x}^{2} + k_{y}^{2})/2m$. The eigenvectors corresponding to the eigenvalues $E_{\pm}$  can be written as
\begin{eqnarray}
\left| \Psi_{+}\right\rangle  = \left(
\begin{array}{c}
\sqrt{\frac{\lambda_{\mathbf{k}} + n_{z}}{2 \lambda_{\mathbf{k}}}} \\
-\frac{n_{x} + i n_{y}}{\sqrt{2 \lambda_{\mathbf{k}} (\lambda_{\mathbf{k}} + n_{z})}} \\
\end{array}
\right),\\
\left| \Psi_{-}\right\rangle  = \left(
\begin{array}{c}
-\frac{n_{x} - i n_{y}}{\sqrt{2 \lambda_{\mathbf{k}} (\lambda_{\mathbf{k}} + n_{z})}} \\
\sqrt{\frac{\lambda_{\mathbf{k}} + n_{z}}{2 \lambda_{\mathbf{k}}}} \\
\end{array}
\right).
\end{eqnarray}
The Berry curvature of the $n$-th ($n=1,2$) band, $\vec{\mathcal{B}}_{n}(\mathbf{k})$, is defined as the rotation of the  Berry connection $\vec{\mathcal{A}}_{n}(\mathbf{k}) = i \left\langle \Psi_{n}\right| \nabla_{\mathbf{k}} \left|\Psi_{n} \right\rangle $ (for details see Refs \onlinecite{Volovik,DiXiaoRevModPhys,NagaosaRevModPhys}). Thus one can write
\begin{equation}
\mathcal{B}_{n}^{z}(\mathbf{k}) = i \left[\frac{\partial}{\partial k_{x}} \left\langle \Psi_{n}\right| \frac{\partial}{\partial k_{y}} \left|\Psi_{n} \right\rangle - \frac{\partial}{\partial k_{y}} \left\langle \Psi_{n}\right| \frac{\partial}{\partial k_{x}} \left|\Psi_{n} \right\rangle \right] .
\end{equation}
Combining Eqs. (28) to (30) we find for the Berry curvature,
\begin{eqnarray}
\mathcal{B}_{\pm}^{z}(\mathbf{k}) =\mp \frac{\alpha^{2} H_{z}}{2 \lambda_{\mathbf{k}}^{3}} .
\end{eqnarray}
Taking the expression above into account, the Berry phase related terms in the electrically generated spin polarization can be written as
\begin{eqnarray}
S_{x}^{II} = \frac{1}{2} e E_{x} \frac{\hbar}{\alpha} \sum_{n} \int \frac{d^{2} \mathbf{k}}{(2\pi)^{2}} f(E_n) \mathcal{B}_{n}^{z}(\mathbf{k})\hspace{1.3cm}\\
S_{y}^{II} = 0\hspace{5.76cm}\\
S_{z}^{II} = -\frac{1}{2} e E_{x} \hbar \sum_{n} \int \frac{d^{2} \mathbf{k}}{(2\pi)^{2}} f(E_{n}) \frac{\alpha k_{y} + H_{x}}{\alpha H_{z}} \mathcal{B}_{n}^{z}.
\end{eqnarray}
Note, these terms disappear in the absence of exchange field.

\section{Spin-orbit torque}

Due to exchange interaction, the current induced spin polarization exerts a torque $\btau$ on the magnetic moment $\mathbf{M}$.
This torque enters the Landau-Lifshitz-Gilbert equation for magnetic dynamics,
\begin{equation}
  \frac{\partial\mathbf{m }}{\partial t} = -\gamma   \mathbf{m } \times \mathbf{h}_{\rm eff}  + \alpha_g   \mathbf{m} \times \frac{\partial\mathbf{m }}{\partial t}  + \btau ,
\label{Eq:LLG_Bulk}
\end{equation}
where $\mathbf{m }=\mathbf{M }/M$ is a unit vector along magnetic moment $\mathbf{M}$, $\mathbf{h}_{\rm eff}$ is the effective magnetic field which includes external magnetic field, dipolar field, and anisotropy field, $\alpha_g$ is the Gilbert damping factor, and $\gamma$ is the giro-magnetic factor.

To find the torque  $\btau$ we write the coupling energy of the magnetic moment and induced spin polarization as  $=(2J/\hbar )\mathbf{S}\cdot \mathbf{M}=-\mathbf{M}\cdot \mathbf{h}_{\rm so}$, where $\mathbf{h}_{\rm so}$ is defined as
\begin{equation}
\mathbf{h}_{\rm so} = -\frac{2J}{\hbar}\mathbf{S}.
\end{equation}
Taking the above into account, one can write
the torque $\btau$ as a sum of a field-like  torque $\btau_f$ and damping-like torque $\btau_d$,
\begin{equation}
\btau=\btau_{f} + \btau_{d}.
\end{equation}
These components can be written in terms of the spin-orbit field  $\mathbf{h}_{\rm so}$ as
\begin{equation}
\btau_{f} = -\gamma   \mathbf{m } \times \mathbf{h}_{\rm so}
\end{equation}
for the field-like term, and
\begin{equation}
\btau_{d} = -\alpha_g \, \gamma \mathbf{m } \times (\mathbf{m }\times\mathbf{h}_{\rm so})
\end{equation}
for the damping-like term.
Since the spin polarization includes terms related to the Berry curvature, the resulting spin-orbit torques include terms related to the Berry curvature as well.

\section{Summary and conclusions}

Using the Matsubara Green function method we have calculated current-induced spin polarization in a magnetized two-dimensional electron gas with the Rashba spin-orbit interaction. The exchange field is shown to have a significant impact on the spin polarization. First, For some orientations of the exchange field, the component of spin polarization that appears in the absence of exchange field can be enhanced by the exchange field, while for other orientations this component can be suppressed. Second, exchange field  also generates the components of spin polarization which are absent in the limit of vanishing exchange field. We also note, that the states at the band edges may become localized due to disorder and the results may be not valid in the localization regime. 

Analytical and/or numerical results have been presented in some special cases, when exchange field is oriented along current or perpendicular to current (in-plane and perpendicular to the plane of 2DEG in the latter case). Numerical results have been also presented in a general case of arbitrary orientation of exchange field. We have found that the exchange field leads to terms in the spin polarization that can be related to the Berry curvature of the corresponding electron bands. Since the calculated spin polarization generates a torque which may induce dynamics of the magnetic moment, this torque includes terms related to the Berry curvature as well.

\begin{acknowledgments}
	This work was supported by  the Polish Ministry
	of Science and Higher Education through a research
	project Iuventus Plus in years 2015-2017 (project
	No. 0083/IP3/2015/73). A.D. also acknowledges support
	from the Fundation for Polish Science (FNP). V.D. acknowledges support from the National
Science Center in Poland under Grant
No. DEC-2012/06/M/ST3/00042. 
\end{acknowledgments}

\appendix
\begin{widetext}

\section{Derivation of Eqs. (\ref{S_x}), (\ref{S_y}), (\ref{S_z})}

The current induced spin polarizaton is evaluated starting from the equation (\ref{SalphaE_final}) that we rewrite in the following form:
\begin{equation}
S_{\alpha} = - e E_{x} \frac{\hbar}{\omega} \int \frac{d^{2} \mathbf{k}}{(2\pi)^{2}} \Bigl(\mathcal{T}_{S_{\alpha}}^{(1)} + \mathcal{T}_{S_{\alpha}}^{(2)}\Bigr),
\end{equation}
where:
\begin{eqnarray}
\label{Tsalpha1}
\mathcal{T}_{S_{\alpha}}^{(1)} = \int \frac{d \varepsilon}{2 \pi} f(\varepsilon) \mathcal{I}_{S_{\alpha}}^{(1)}(\varepsilon + \omega, \varepsilon),\\
\label{Tsalpha2}
\mathcal{T}_{S_{\alpha}}^{(2)} = \int \frac{d \varepsilon}{2 \pi} f(\varepsilon) \mathcal{I}_{S_{\alpha}}^{(2)}(\varepsilon, \varepsilon - \omega),
\end{eqnarray}
and the following notation has been introduced:
\begin{eqnarray}
\mathcal{I}_{S_{\alpha}}^{(1)}(\varepsilon + \omega, \varepsilon) = {\mathrm{Tr}} \left\{ \hat{s}_{\alpha} G_{\mathbf{k}}^{R}(\varepsilon + \omega) \hat{v}_{x} [G_{\mathbf{k}}^{R}(\varepsilon) - G_{\mathbf{k}}^{A}(\varepsilon)]\right\},\hspace{0.8cm}\\
\mathcal{I}_{S_{\alpha}}^{(2)}(\varepsilon, \varepsilon - \omega) = {\mathrm{Tr}} \left\{\hat{s}_{\alpha} [G_{\mathbf{k}}^{R}(\varepsilon) - G_{\mathbf{k}}^{A}(\varepsilon)] \hat{v}_{x} G_{\mathbf{k}}^{A}(\varepsilon - \omega) \right\},\hspace{0.8cm}
\end{eqnarray}

According to the above notation the $S_{x}$ component of spin polarization is described by the following expressions:
\begin{eqnarray}
\label{Isx1}
\mathcal{I}_{S_{x}}^{(1)}(\varepsilon + \omega, \varepsilon) = \frac{\hbar^{2} k_{x}}{2 m \lambda_{\mathbf{k}}} (\alpha k_{y} - H_{x})
\Bigl[G_{\mathbf{k} -}^{R}(\varepsilon+ \omega) G_{\mathbf{k} -}^{A}(\varepsilon) - G_{\mathbf{k} -}^{R}(\varepsilon + \omega) G_{\mathbf{k} -}^{R}(\varepsilon)
+ G_{\mathbf{k} +}^{R}(\varepsilon + \omega) G_{\mathbf{k} +}^{R} (\varepsilon) - G_{\mathbf{k} +}^{R}(\varepsilon + \omega) G_{\mathbf{k} +}^{A} (\varepsilon) \Bigr]\nonumber\\
- \frac{\alpha}{2 \lambda_{\mathbf{k}}^{2}} (\alpha k_{y} - H_{x}) (\alpha k_{x} + H_{y})
\Bigl[ G_{\mathbf{k} -}^{R}(\varepsilon + \omega) G_{\mathbf{k} -}^{A}(\varepsilon) - G_{\mathbf{k} -}^{R}(\varepsilon + \omega) G_{\mathbf{k} +}^{A}(\varepsilon)
-G_{\mathbf{k} -}^{R}(\varepsilon + \omega) G_{\mathbf{k} -}^{R}(\varepsilon) + G_{\mathbf{k} -}^{R}(\varepsilon + \omega) G_{\mathbf{k} +}^{R}(\varepsilon)\nonumber\\
- G_{\mathbf{k} +}^{R}(\varepsilon + \omega) G_{\mathbf{k} -}^{A}(\varepsilon) + G_{\mathbf{k} +}^{R}(\varepsilon + \omega) G_{\mathbf{k} +}^{A}(\varepsilon)
+ \Bigl. G_{\mathbf{k} +}^{R}(\varepsilon + \omega) G_{\mathbf{k} -}^{R}(\varepsilon) - G_{\mathbf{k} +}^{R}(\varepsilon + \omega) G_{\mathbf{k} +}^{R}(\varepsilon)\Bigr]\nonumber\\
- i \frac{\alpha}{2\lambda_{\mathbf{k}}} H_{z}
\Bigl[ G_{\mathbf{k} -}^{R}(\varepsilon + \omega) G_{\mathbf{k} +}^{A}(\varepsilon) - G_{\mathbf{k} -}^{R}(\varepsilon + \omega) G_{\mathbf{k} +}^{R}(\varepsilon)
 -G_{\mathbf{k} +}^{R}(\varepsilon + \omega) G_{\mathbf{k} -}^{A}(\varepsilon) + G_{\mathbf{k} +}^{R}(\varepsilon + \omega) G_{\mathbf{k} -}^{R}(\varepsilon)\Bigr]\nonumber\\
\end{eqnarray}
\begin{eqnarray}
\label{Isx2}
\mathcal{I}_{S_{x}}^{(2)}(\varepsilon, \varepsilon - \omega) = \frac{\hbar^{2} k_{x}}{2 m \lambda_{\mathbf{k}}} (\alpha k_{y} - H_{x})
\Bigl[ G_{\mathbf{k} -}^{A}(\varepsilon) G_{\mathbf{k} -}^{A}(\varepsilon - \omega) - G_{\mathbf{k} +}^{A}(\varepsilon) G_{\mathbf{k} +}^{A}(\varepsilon - \omega)
-G_{\mathbf{k} -}^{R}(\varepsilon) G_{\mathbf{k} -}^{A}(\varepsilon) + G_{\mathbf{k} +}^{R}(\varepsilon) G_{\mathbf{k} +}^{A}(\varepsilon - \omega) \Bigr]\nonumber\\
- \frac{\alpha}{2 \lambda_{\mathbf{k}}^{2}} (\alpha k_{x} + H_{y}) (\alpha k_{y} - H_{x})
\Bigl[ G_{\mathbf{k} -}^{A}(\varepsilon) G_{\mathbf{k} -}^{A}(\varepsilon - \omega) - G_{\mathbf{k} +}^{A}(\varepsilon) G_{\mathbf{k} -}^{A}(\varepsilon - \omega) \Bigr.
- G_{\mathbf{k} -}^{R}(\varepsilon) G_{\mathbf{k} -}^{A}(\varepsilon - \omega) + G_{\mathbf{k} +}^{R}(\varepsilon) G_{\mathbf{k} -}^{A}(\varepsilon - \omega)\nonumber\\
- G_{\mathbf{k} -}^{A}(\varepsilon) G_{\mathbf{k} +}^{A}(\varepsilon - \omega) + G_{\mathbf{k} +}^{A}(\varepsilon) G_{\mathbf{k} +}^{A}(\varepsilon - \omega)
+ \Bigl. G_{\mathbf{k} -}^{R}(\varepsilon) G_{\mathbf{k} +}^{A}(\varepsilon - \omega) - G_{\mathbf{k} +}^{R}(\varepsilon) G_{\mathbf{k} +}^{A}(\varepsilon - \omega)\Bigr]\nonumber\\
- i \frac{\alpha}{\lambda_{2\mathbf{k}}} H_{z} \Bigl[ G_{\mathbf{k} -}^{A}(\varepsilon) G_{\mathbf{k} +}^{A}(\varepsilon - \omega) - G_{\mathbf{k} +}^{A}(\varepsilon) G_{\mathbf{k} -}^{A}(\varepsilon - \omega) 
-G_{\mathbf{k} -}^{R}(\varepsilon) G_{\mathbf{k} +}^{A}(\varepsilon - \omega) + G_{\mathbf{k} +}^{R}(\varepsilon) G_{\mathbf{k} -}^{A}(\varepsilon - \omega) \Bigr]\nonumber\\
\end{eqnarray}
Inserting Eqs. (\ref{Isx1}), (\ref{Isx2}) into Eqs. (\ref{Tsalpha1}) and (\ref{Tsalpha2}) respectively we get:

\begin{eqnarray}
\mathfrak{Re}\Bigl[ \mathcal{T}_{S_{x}}^{(1)} + \mathcal{T}_{S_{x}}^{(2)}\Bigr]
=\omega \frac{\hbar^{2} k_{x}}{2 m \lambda_{\mathbf{k}}} (\alpha k_{y} - H_{x}) \frac{2 \Gamma}{\omega^{2} + (2\Gamma)^{2}} [f'(E_{+}) - f'(E_{-})]
\hspace{7.5cm}\nonumber\\
+ \omega \frac{\alpha}{2 \lambda_{\mathbf{k}}^{2}} (\alpha k_{x} + H_{y}) (\alpha k_{y} - H_{x}) \frac{2 \Gamma}{\omega^{2} + (2\Gamma)^{2}} [f'(E_{+}) + f'(E_{-})]\hspace{7.2cm}
\nonumber\\
- \omega \Gamma \frac{\alpha}{ \lambda_{\mathbf{k}}^{2}} (\alpha k_{x} + H_{y}) (\alpha k_{y} - H_{x})
\left( \frac{f'(E_{-})}{(E_{+} - E_{-} - \omega)^{2} + (2\Gamma)^{2}} +  \frac{f'(E_{+})}{(E_{+} - E_{-} + \omega)^{2} + (2\Gamma)^{2}} \right)\hspace{2.8cm} \nonumber\\
-\omega \frac{\alpha}{2 \lambda_{\mathbf{k}}} H_{z}
\left( \frac{E_{+} - E_{-} - \omega}{(E_{+} - E_{-} - \omega)^{2} + (2\Gamma)^{2}}f'(E_{-}) -  \frac{E_{+} - E_{-} + \omega}{(E_{+} - E_{-} + \omega)^{2} + (2\Gamma)^{2}} f'(E_{+}) \right)\hspace{3.9cm}\nonumber\\
+ \omega \frac{\alpha}{2 \lambda_{\mathbf{k}}} H_{z} \frac{E_{+} - E_{-}}{(E_{+} - E_{-})^{2} - \omega^{2}} [f'(E_{+}) + f'(E_{-})]
- \omega \frac{\alpha}{\lambda_{\mathbf{k}}} H_{z} \frac{f(E_{+}) - f(E_{-})}{(E_{+} - E_{-})^{2} - \omega^{2}}+ \omega^{2} \frac{\alpha}{2 \lambda_{\mathbf{k}}} H_{z}\frac{f'(E_{-}) - f'(E_{+})}{(E_{+} - E_{-})^{2} - \omega^{2}}\hspace{1.5cm}
\end{eqnarray}

In the limit of $\omega \rightarrow 0$ we find $x$ component of current-induced spin polarization given by Eq.(\ref{S_x}).

In turn, the $S_{y}$ component of spin polarization is expressed by the following functions:
\begin{eqnarray}\label{Isy1}
\mathcal{I}_{S_{y}}^{(1)}(\varepsilon + \omega, \varepsilon) = \frac{\alpha}{2} \Bigl[ G_{\mathbf{k} -}^{R}(\varepsilon+\omega) G_{\mathbf{k} -}^{A}(\varepsilon) + G_{\mathbf{k} +}^{R}(\varepsilon+\omega) G_{\mathbf{k} +}^{A}(\varepsilon)\Bigr.
- \Bigl. G_{\mathbf{k} -}^{R}(\varepsilon+\omega) G_{\mathbf{k} -}^{R}(\varepsilon) - G_{\mathbf{k} +}^{R}(\varepsilon+\omega) G_{\mathbf{k} +}^{R}(\varepsilon)\Bigr]\hspace{2cm}
\nonumber\\
- \frac{\hbar^{2} k_{x}}{2 m \lambda_{\mathbf{k}}} (\alpha k_{x} + H_{y})
\Bigl[ G_{\mathbf{k} -}^{R}(\varepsilon+\omega) G_{\mathbf{k} -}^{A}(\varepsilon) - G_{\mathbf{k} -}^{R}(\varepsilon+\omega) G_{\mathbf{k} -}^{R}(\varepsilon)\Bigr.
- \Bigl. G_{\mathbf{k} +}^{R}(\varepsilon+\omega) G_{\mathbf{k} +}^{A}(\varepsilon) + G_{\mathbf{k} +}^{R}(\varepsilon+\omega) G_{\mathbf{k} +}^{R}(\varepsilon)\Bigr]\nonumber\\
- \frac{\alpha}{2 \lambda_{\mathbf{k}}^{2}} \left[ (\alpha k_{y} - H_{x})^{2} + H_{z}^{2}\right]
\Bigl[ G_{\mathbf{k} -}^{R}(\varepsilon+\omega) G_{\mathbf{k} -}^{A}(\varepsilon) - G_{\mathbf{k} +}^{R}(\varepsilon+\omega) G_{\mathbf{k} -}^{A}(\varepsilon)\Bigr.
-  G_{\mathbf{k} -}^{R}(\varepsilon+\omega) G_{\mathbf{k} -}^{R}(\varepsilon) + G_{\mathbf{k} -}^{R}(\varepsilon+\omega) G_{\mathbf{k} +}^{R}(\varepsilon)\nonumber\\
+ G_{\mathbf{k} -}^{R}(\varepsilon+\omega) G_{\mathbf{k} +}^{R}(\varepsilon) - G_{\mathbf{k} +}^{R}(\varepsilon+\omega) G_{\mathbf{k} +}^{R}(\varepsilon)
- \Bigl. G_{\mathbf{k} -}^{R}(\varepsilon+\omega) G_{\mathbf{k} +}^{A}(\varepsilon) + G_{\mathbf{k} +}^{R}(\varepsilon+\omega) G_{\mathbf{k} +}^{A}(\varepsilon)\Bigr]\nonumber\\
\end{eqnarray}

\begin{eqnarray}\label{Isy2}
\mathcal{I}_{S_{y}}^{(2)}(\varepsilon, \varepsilon - \omega) = \frac{\alpha}{2} \Bigl[ G_{\mathbf{k} -}^{A}(\varepsilon) G_{\mathbf{k} -}^{A}(\varepsilon - \omega) - G_{\mathbf{k} -}^{R}(\varepsilon) G_{\mathbf{k} -}^{A}(\varepsilon - \omega) \Bigr.
+ \Bigl. G_{\mathbf{k} +}^{A}(\varepsilon) G_{\mathbf{k} +}^{A}(\varepsilon - \omega) - G_{\mathbf{k} +}^{R}(\varepsilon) G_{\mathbf{k} +}^{A}(\varepsilon - \omega) \Bigr]\hspace{2cm}\nonumber\\
- \frac{\hbar^{2} k_{x}}{2 m \lambda_{\mathbf{k}}} (\alpha k_{x} + H_{y})
\Bigl[ G_{\mathbf{k} -}^{A}(\varepsilon) G_{\mathbf{k} -}^{A}(\varepsilon-\omega) - G_{\mathbf{k} +}^{A}(\varepsilon) G_{\mathbf{k} +}^{A}(\varepsilon-\omega)\Bigr.
- \Bigl. G_{\mathbf{k} -}^{R}(\varepsilon) G_{\mathbf{k} -}^{A}(\varepsilon-\omega) + G_{\mathbf{k} +}^{R}(\varepsilon) G_{\mathbf{k} +}^{A}(\varepsilon-\omega)  \Bigr]\nonumber\\
- \frac{\alpha}{2 \lambda_{\mathbf{k}}^{2}} \left[ (\alpha k_{y} - H_{x})^{2} + H_{z}^{2}\right]
\Bigl[ G_{\mathbf{k} -}^{A}(\varepsilon) G_{\mathbf{k} -}^{A}(\varepsilon-\omega) - G_{\mathbf{k} +}^{A}(\varepsilon) G_{\mathbf{k} -}^{A}(\varepsilon-\omega)\Bigr.
- G_{\mathbf{k} -}^{R}(\varepsilon) G_{\mathbf{k} -}^{A}(\varepsilon-\omega) + G_{\mathbf{k} +}^{R}(\varepsilon) G_{\mathbf{k} -}^{A}(\varepsilon-\omega)\nonumber\\
- G_{\mathbf{k} -}^{A}(\varepsilon) G_{\mathbf{k} +}^{A}(\varepsilon-\omega)+ G_{\mathbf{k} +}^{A}(\varepsilon) G_{\mathbf{k} +}^{A}(\varepsilon-\omega)
+ \Bigl. G_{\mathbf{k} -}^{R}(\varepsilon) G_{\mathbf{k} +}^{A}(\varepsilon-\omega) - G_{\mathbf{k} +}^{R}(\varepsilon) G_{\mathbf{k} +}^{A}(\varepsilon-\omega)\Bigr]
\nonumber\\
\end{eqnarray}
After integration over $\varepsilon$  in Eqs. (\ref{Tsalpha1}) and (\ref{Tsalpha2}) with integrands given by  (\ref{Isy1}),(\ref{Isy2}) we obtain the following expression:

\begin{eqnarray}
\mathfrak{Re}\Bigl[ \mathcal{T}_{S_{y}}^{(1)} + \mathcal{T}_{S_{y}}^{(2)}\Bigr]
=
- \frac{\hbar^{2} k_{x}}{2 m \lambda_{\mathbf{k}}} (\alpha k_{x} + H_{y}) \frac{2 \Gamma \omega}{\omega^{2} + (2 \Gamma)^{2}} [f'(E_{+}) - f'(E_{-})]
- \frac{\alpha}{ \lambda_{\mathbf{k}}^{2}}(\alpha k_{x} + H_{y})^{2}
\frac{2 \Gamma \omega}{\omega^{2} + (2 \Gamma)^{2}} [f'(E_{+}) - f'(E_{-})]
\nonumber\\
- \frac{\alpha}{\lambda_{\mathbf{k}}^{2}} \Gamma \omega \left[ (\alpha k_{y} - H_{x})^{2} + H_{z}^{2} \right]
 %
\left(\frac{f'(E_{-}) }{(E_{+} - E_{-} - \omega)^{2} + (2\Gamma)^{2}} + \frac{f'(E_{+}) }{(E_{+} - E_{-} + \omega)^{2} + (2\Gamma)^{2}}\right)\hspace{2cm}
\end{eqnarray}
In the limit $\omega \rightarrow 0$ we obtain the formula describing $y$ component of current-induced spin polarization given by Eq.(\ref{S_y}).

Finally, the $S_{z}$ component of the nonequilibrium spin polarization is described by following traces:
\begin{eqnarray}
\mathcal{I}_{S_{z}}^{(1)}(\varepsilon + \omega, \varepsilon) = \frac{\alpha^{2} k_{x}}{2 \lambda_{\mathbf{k}}^{2}} H_{z}
\Bigl[ G_{\mathbf{k} -}^{R}(\varepsilon + \omega) G_{\mathbf{k} -}^{A}(\varepsilon) - G_{\mathbf{k} -}^{R}(\varepsilon + \omega) G_{\mathbf{k} +}^{A}(\varepsilon) \Bigr.
- G_{\mathbf{k} -}^{R}(\varepsilon + \omega) G_{\mathbf{k} -}^{R}(\varepsilon) + G_{\mathbf{k} -}^{R}(\varepsilon + \omega) G_{\mathbf{k} +}^{R}(\varepsilon)\hspace{1.2cm}\nonumber\\
- G_{\mathbf{k} +}^{R}(\varepsilon + \omega) G_{\mathbf{k} -}^{A}(\varepsilon) + G_{\mathbf{k} +}^{R}(\varepsilon + \omega) G_{\mathbf{k} +}^{A}(\varepsilon)
+ \Bigl. G_{\mathbf{k} +}^{R}(\varepsilon + \omega) G_{\mathbf{k} -}^{R}(\varepsilon) - G_{\mathbf{k} +}^{R}(\varepsilon + \omega) G_{\mathbf{k} +}^{R}(\varepsilon)\Bigr]\nonumber\\
+ \frac{\hbar^{2} k_{x}}{2 m \lambda_{\mathbf{k}}} H_{z} 
\Bigl[ G_{\mathbf{k} -}^{R}(\varepsilon + \omega) G_{\mathbf{k} -}^{R}(\varepsilon) - G_{\mathbf{k} -}^{R}(\varepsilon + \omega) G_{\mathbf{k} -}^{A}(\varepsilon)\Bigr.
+ \Bigl. G_{\mathbf{k} +}^{R}(\varepsilon + \omega) G_{\mathbf{k} +}^{A}(\varepsilon) - G_{\mathbf{k} +}^{R}(\varepsilon + \omega) G_{\mathbf{k} +}^{R}(\varepsilon)\Bigr]\hspace{1.0cm}\nonumber\\
+ \frac{\alpha}{2 \lambda_{\mathbf{k}}^{2}} H_{y} H_{z}
\Bigl[ G_{\mathbf{k} -}^{R}(\varepsilon + \omega) G_{\mathbf{k} -}^{A}(\varepsilon) - G_{\mathbf{k} -}^{R}(\varepsilon + \omega) G_{\mathbf{k} +}^{A}(\varepsilon)\Bigr.
- G_{\mathbf{k} -}^{R}(\varepsilon + \omega) G_{\mathbf{k} -}^{R}(\varepsilon) + G_{\mathbf{k} -}^{R}(\varepsilon + \omega) G_{\mathbf{k} +}^{R}(\varepsilon)\hspace{1.0cm}\nonumber\\
- G_{\mathbf{k} +}^{R}(\varepsilon + \omega) G_{\mathbf{k} -}^{A}(\varepsilon) + G_{\mathbf{k} +}^{R}(\varepsilon + \omega) G_{\mathbf{k} +}^{A}(\varepsilon)
+ \Bigl. G_{\mathbf{k} +}^{R}(\varepsilon + \omega) G_{\mathbf{k} -}^{R} (\varepsilon) - G_{\mathbf{k} +}^{R}(\varepsilon + \omega) G_{\mathbf{k} +}^{R}(\varepsilon) \Bigr]\nonumber\\
+ i \frac{\alpha}{2 \lambda_{\mathbf{k}}} (H_{x} - \alpha k_{y})
\Bigl[ G_{\mathbf{k} -}^{R}(\varepsilon + \omega) G_{\mathbf{k} +}^{A}(\varepsilon) - G_{\mathbf{k} -}^{R}(\varepsilon + \omega) G_{\mathbf{k} +}^{R}(\varepsilon)\Bigr.
- \Bigl. G_{\mathbf{k} +}^{R}(\varepsilon + \omega) G_{\mathbf{k} -}^{A}(\varepsilon) + G_{\mathbf{k} +}^{R}(\varepsilon + \omega) G_{\mathbf{k} -}^{R}(\varepsilon) \Bigr]\nonumber\\
\end{eqnarray}

\begin{eqnarray}
\mathcal{I}_{S_{z}}^{(2)}(\varepsilon, \varepsilon - \omega) = \frac{\alpha^{2} k_{x}}{2 \lambda_{\mathbf{k}}^{2}} H_{z}
\Bigl[ G_{\mathbf{k} -}^{A}(\varepsilon) G_{\mathbf{k} -}^{A}(\varepsilon - \omega) - G_{\mathbf{k} +}^{A}(\varepsilon) G_{\mathbf{k} -}^{A}(\varepsilon - \omega)\Bigr.
- G_{\mathbf{k} -}^{R}(\varepsilon) G_{\mathbf{k} -}^{A}(\varepsilon - \omega) + G_{\mathbf{k} +}^{R}(\varepsilon) G_{\mathbf{k} -}^{A}(\varepsilon - \omega)\hspace{1.5cm}\nonumber\\
- G_{\mathbf{k} -}^{A}(\varepsilon) G_{\mathbf{k} +}^{A}(\varepsilon - \omega) + G_{\mathbf{k} +}^{A}(\varepsilon) G_{\mathbf{k} +}^{A}(\varepsilon - \omega)
+ \Bigl. G_{\mathbf{k} -}^{R}(\varepsilon) G_{\mathbf{k} +}^{A}(\varepsilon - \omega) - G_{\mathbf{k} +}^{R}(\varepsilon) G_{\mathbf{k} +}^{A}(\varepsilon - \omega)\Bigr]\nonumber\\
+ \frac{\hbar^{2} k_{x}}{2 m \lambda_{\mathbf{k}}} H_{z} 
\Bigl[ G_{\mathbf{k} +}^{A}(\varepsilon) G_{\mathbf{k} +}^{A}(\varepsilon - \omega) - G_{\mathbf{k} -}^{A}(\varepsilon) G_{\mathbf{k} -}^{A}(\varepsilon - \omega) \Bigr.
+ \Bigl. G_{\mathbf{k} -}^{R}(\varepsilon) G_{\mathbf{k} -}^{A}(\varepsilon - \omega) - G_{\mathbf{k} +}^{R}(\varepsilon) G_{\mathbf{k} +}^{A}(\varepsilon - \omega)\Bigr]\hspace{1.1cm}\nonumber\\
+ \frac{\alpha}{2 \lambda_{\mathbf{k}}^{2}} H_{y} H_{z}
\Bigl[ G_{\mathbf{k} -}^{A}(\varepsilon) G_{\mathbf{k} -}^{A}(\varepsilon - \omega) - G_{\mathbf{k} +}^{A}(\varepsilon) G_{\mathbf{k} -}^{A}(\varepsilon - \omega) \Bigr.
- G_{\mathbf{k} -}^{R}(\varepsilon) G_{\mathbf{k} -}^{A}(\varepsilon - \omega) + G_{\mathbf{k} +}^{R}(\varepsilon) G_{\mathbf{k} -}^{A}(\varepsilon - \omega)\hspace{1.1cm}\nonumber\\
- G_{\mathbf{k} -}^{A}(\varepsilon) G_{\mathbf{k} +}^{A}(\varepsilon - \omega) + G_{\mathbf{k} +}^{A}(\varepsilon) G_{\mathbf{k} +}^{A}(\varepsilon - \omega)
+ \Bigl. G_{\mathbf{k} -}^{R}(\varepsilon) G_{\mathbf{k} +}^{A}(\varepsilon - \omega) - G_{\mathbf{k} +}^{R}(\varepsilon) G_{\mathbf{k} +}^{A}(\varepsilon - \omega)\Bigr]\nonumber\\
+ i \frac{\alpha}{2 \lambda_{\mathbf{k}}} (H_{x} - \alpha k_{y})
\Bigl[ G_{\mathbf{k} -}^{A}(\varepsilon) G_{\mathbf{k} +}^{A}(\varepsilon - \omega) - G_{\mathbf{k} +}^{A}(\varepsilon) G_{\mathbf{k} -}^{A}(\varepsilon - \omega)\Bigr.
- \Bigl. G_{\mathbf{k} -}^{R}(\varepsilon) G_{\mathbf{k} +}^{A}(\varepsilon - \omega)  + G_{\mathbf{k} +}^{R}(\varepsilon) G_{\mathbf{k} -}^{A}(\varepsilon - \omega)\Bigr]\nonumber\\
\end{eqnarray}
These two equations combining with Eqs. (\ref{Tsalpha1}) and (\ref{Tsalpha2}) lead to the following expression:
\begin{eqnarray}
\mathfrak{Re}\Bigl[ \mathcal{T}_{S_{z}}^{I} + \mathcal{T}_{S_{z}}^{II}\Bigr]
= - \omega \frac{2 \Gamma}{\omega^{2} + (2\Gamma)^{2}} \frac{\hbar^{2} k_{x}}{2 m \lambda_{\mathbf{k}}} H_{z} [f'(E_{+}) - f'(E_{-})]
- \omega \frac{\alpha}{2 \lambda_{\mathbf{k}}^{2}} H_{z} (\alpha k_{x} + H_{y})
 \frac{2 \Gamma}{\omega^{2} + (2\Gamma)^{2}} [f'(E_{+}) + f'(E_{-})]\nonumber\\
 + \omega \Gamma \frac{\alpha}{\lambda_{\mathbf{k}}^{2}} (\alpha k_{x} + H_{y}) H_{z} \Big[\frac{f'(E_{-})}{(E_{+} - E_{-} - \omega)^{2} + (2\Gamma)^{2}} + \frac{f'(E_{+})}{(E_{+} - E_{-} + \omega)^{2} + (2\Gamma)^{2}} \Big]\nonumber\\
 + \omega \frac{\alpha}{2 \lambda_{\mathbf{k}}} (H_{x} - \alpha k_{y}) \Big[\frac{E_{+} - E_{-} - \omega}{(E_{+} - E_{-} - \omega)^{2} + (2\Gamma)^{2}} f'(E_{-}) + \frac{E_{+} - E_{-} + \omega}{(E_{+} - E_{-} + \omega)^{2} + (2\Gamma)^{2}} f'(E_{+})  \Big]\nonumber\\
 - \omega \alpha (H_{x} - \alpha k_{y}) \frac{f'(E_{+}) + f'(E_{-})}{(E_{+} - E_{-})^{2} - \omega^{2}} - \omega \frac{\alpha}{\lambda_{\mathbf{k}}} (H_{x} - \alpha k_{y}) \frac{f(E_{-}) - f(E_{+})}{(E_{+} - E_{-})^{2} - \omega^{2}} - \omega^{2} \frac{\alpha}{2 \lambda_{\mathbf{k}}} (H_{x} - \alpha k_{y}) \frac{f'(E_{-}) - f'(E_{+})}{(E_{+} - E_{-})^{2} - \omega^{2}}\nonumber\\
\end{eqnarray}
In the dc-limit we get Eq. (\ref{S_z}).

\end{widetext}


\end{document}